\documentclass[]{spie}  
\pdfoutput=1
\usepackage[]{graphicx}
\usepackage{booktabs}
\usepackage{amsmath, amssymb}
\usepackage{subfigure}

\title{CHARIS Science:  Performance Simulations for the Subaru Telescope's
Third-Generation of Exoplanet Imaging Instrumentation}
\author{Timothy D. Brandt\supit{a}, Michael W. McElwain\supit{b}, Markus 
Janson\supit{c}, Gillian R. Knapp\supit{d}, Kyle Mede\supit{e}, Mary Anne Limbach\supit{d},
Tyler Groff\supit{d}, Adam Burrows\supit{d}, 
James E. Gunn\supit{d},
Olivier Guyon\supit{f,g}, Jun Hashimoto\supit{h}, Masahiko Hayashi\supit{i},
Nemanja Jovanovic\supit{f},
N. Jeremy Kasdin\supit{d}, Masayuki Kuzuhara\supit{i,j}, 
Robert H. Lupton\supit{d}, Frantz 
Martinache\supit{f,k}, Satoko Sorahana\supit{l}, David S.~Spiegel\supit{a}, Naruhisa Takato\supit{f},
Motohide Tamura\supit{i,e}, Edwin L. Turner\supit{d,m}, Robert 
Vanderbei\supit{d}, John Wisniewski\supit{h}
\skiplinehalf
\supit{a}Institute for Advanced Study, Princeton, NJ, USA;\\
\supit{b}Goddard Space Flight Center, Greenbelt, MD, USA; \\
\supit{c}Queens University Belfast, Belfast, Northern Ireland, UK; \\
\supit{d}Princeton University, Princeton, NJ, USA; \\
\supit{e}University of Tokyo, Tokyo, Japan; \\
\supit{f}Subaru Headquarters, National Astronomical Observatory of Japan,
Hilo, HI, USA;\\
\supit{g}University of Arizona, Tucson, AZ, USA; \\
\supit{h}University of Oklahoma, Norman, OK, USA; \\
\supit{i}National Astronomical Observatory of Japan, Tokyo, Japan;\\
\supit{j}Tokyo Institute of Technology, Tokyo, Japan; \\
\supit{k}Laboratoire Lagrange, Observatoire de la C\^ote d'Azur, Nice, France; \\
\supit{l}Nagoya University, Nagoya, Japan; \\
\supit{m}Kavli IPMU (WPI), University of Tokyo, Tokyo, Japan.
}

\authorinfo{Further author information: Send correspondence to 
Timothy Brandt)\\email: tbrandt@ias.edu}
 
\begin{document} 
\maketitle 

\begin{abstract}
We describe the expected scientific capabilities of CHARIS, a high-contrast integral-field spectrograph (IFS) currently under construction for the Subaru telescope.  CHARIS is part of a new generation of instruments, enabled by extreme adaptive optics (AO) systems (including SCExAO at Subaru), that promise greatly improved contrasts at small angular separation thanks to their ability to use spectral information to distinguish planets from quasistatic speckles in the stellar point-spread function (PSF).  CHARIS is similar in concept to GPI and SPHERE, on Gemini South and the Very Large Telescope, respectively, but will be unique in its ability to simultaneously cover the entire near-infrared $J$, $H$, and $K$ bands with a low-resolution mode.  This extraordinarily broad wavelength coverage will enable spectral differential imaging down to angular separations of a few $\lambda/D$, corresponding to $\sim$$0.\!\!''1$.  SCExAO will also offer contrast approaching $10^{-5}$ at similar separations, $\sim$$0.\!\!''1$--$0.\!\!''2$.  The discovery yield of a CHARIS survey will depend on the exoplanet distribution function at around 10 AU.  If the distribution of planets discovered by radial velocity surveys extends unchanged to $\sim$20 AU, observations of $\sim$200 mostly young, nearby stars targeted by existing high-contrast instruments might find $\sim$1--3 planets.  Carefully optimizing the target sample could improve this yield by a factor of a few, while an upturn in frequency at a few AU could also increase the number of detections.  CHARIS, with a higher spectral resolution mode of $R \sim 75$, will also be among the best instruments to characterize planets and brown dwarfs like HR 8799 cde and $\kappa$ And b.  
\end{abstract}


\section{INTRODUCTION}
\label{sec:intro} 

Thousands of exoplanets have recently been discovered around nearby stars\cite{}, the overwhelming majority by indirect techniques like transits and radial velocity variations in their host stars.  These techniques become ineffective at large separations: the amplitude of the radial velocity signal drops as the period increases, while the probability and frequency of transits both fall precipitously.  With adaptive optics (AO) systems on 8--10 meter telescopes, it has recently become possible to directly image massive planets shining by their residual heat of formation.  

Direct imaging is sensitive to a different planet population than the radial velocity and transit methods, and is particularly well-suited to finding more massive brown dwarf companions.  By obtaining light directly from a substellar companion, imaging also offers the opportunity to characterize its effective temperature, gravity, and chemistry, and ultimately to learn about its mode of formation.  Spectra have now been taken of a few planets using direct imaging\cite{Konopacky+Barman+Macintosh+etal_2013, Oppenheimer+Baranec+Beichman+etal_2013}, offering hints of their chemical compositions.  The addition of many more spectra in the coming years will enable the spectral classification of substellar objects extending far below the deuterium-burning limit.  

CHARIS, the Coronagraphic High Angular Resolution Imaging Spectrograph\cite{Peters+Groff+Kasdin+etal_2012, Groff+Kasdin+Peters+etal_2014}, is part of a new generation of high-contrast instruments called integral-field spectrographs (IFSs).  These instruments obtain a spectrum of each spatial element of their fields-of-view, producing a data cube with two spatial ($x$, $y$) and one spectral ($\lambda$) dimension.  The spectral information allows post-processing software to take advantage of the fact that the stellar point-spread function (PSF), as a diffraction phenomenon, scales with wavelength.  Physical companions can thus be separated from quasistatic speckles.  Combined with very high performance AO systems, instruments like GPI\cite{Macintosh+Graham+Palmer+etal_2008} and SPHERE\cite{Beuzit+Feldt+Dohlen+etal_2008} in the Southern Hemisphere are beginning large-scale exoplanet surveys with unprecedented contrast at small angular separations.  

CHARIS is similar in concept to GPI and SPHERE, but lies in the Northern Hemisphere.  CHARIS will offer two dispersion modes, $R \sim 18$ and $R \sim 75$, and will cover the near-infrared from 1.15 to 2.4 $\mu$m (the $J$, $H$, and $K$ bands).  Its simultaneous spectral coverage from $J$ to $K$ (1.15--2.4 $\mu$m) in its low-resolution mode will be unique in its class of instruments.  With a very high-quality input beam from SCExAO\cite{Guyon+Martinache+Garrel+etal_2010}, the Subaru Coronagraphic Extreme Adaptive Optics system, CHARIS will achieve contrasts approaching $10^{-5}$ at an inner working angle of $\sim$$0.\!\!''1$ ($\sim$3 $\lambda/D$), with the contrast improving by at least an order of magnitude at $\sim$1$''$.  In this paper, we review CHARIS's design and expected performance, and summarize its anticipated science yield.

\section{CHARIS Specifications} \label{sec:specifications}

\begin{figure}
\includegraphics[width=\linewidth]{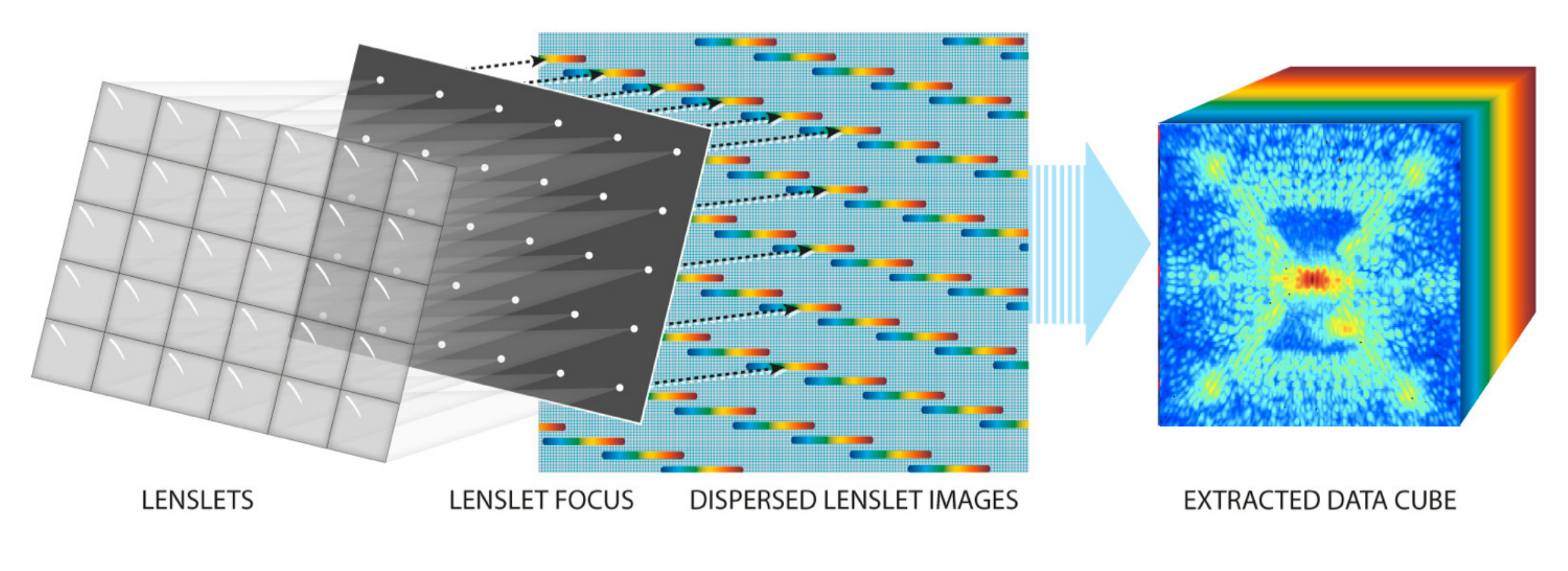}
\caption[example] { \label{fig:datacube} The basic concept of a lenslet-based IFS.  A lenslet array samples the image plane, focusing the light into a sparse grid.  Light from each spatial element is then dispersed by a prism before eventually landing on the detector.  We then use the thousands of mini spectra to reconstruct the data cube, the spectrum of each spatial element of the field-of-view.}
\end{figure}

CHARIS is a lenslet-based IFS that will sit behind SCExAO on the Subaru Telescope's Nasmyth platform.  CHARIS will use its lenslet array to sample the image plane, rendering the image sparse and providing space for a prism to disperse.  It will produce a data cube, with two spatial and one spectral dimension (see Figure \ref{fig:datacube}).  CHARIS's basic design was first used in the TIGER IFS\cite{Bacon+Adam+Baranne+etal_1995}, and is similar to that used in GPI and SPHERE.  

CHARIS adds several enhancements to the basic design shown in Figure \ref{fig:datacube}.  These include a pinhole grid printed onto the back of the lenslet array to reduce diffractive cross-talk between spatial and spectral positions, and multiple dispersing modes to optimize CHARIS for both sensitivity and resolving power.  CHARIS will be the first IFS to use L-BBH, a new glass developed by Ohara with a remarkably flat index of refraction across the near-infrared.  This material allows our high-resolution mode ($R \sim 75$) to cover the $J$, $H$, or $K$ band in a single observation.  GPI, for comparison, must split the $K$ band in half because their prism's dispersion sharply rises redward of $\sim$2 $\mu$m.  

CHARIS's use of the new LBBH glass also allows for its unique feature: a low spectral resolution mode ($R \sim 18$) in which the entire near-infrared from $J$ to $K$ (1.15 to 2.4 $\mu$m) may be observed simultaneously.  This not only allows CHARIS to collect more photons, but provides a full factor of two in wavelength coverage for spectral differential imaging (SDI).  Table \ref{tab:charis_spec} lists CHARIS's main specifications.

\begin{figure}
\centering\includegraphics[width=0.8\linewidth]{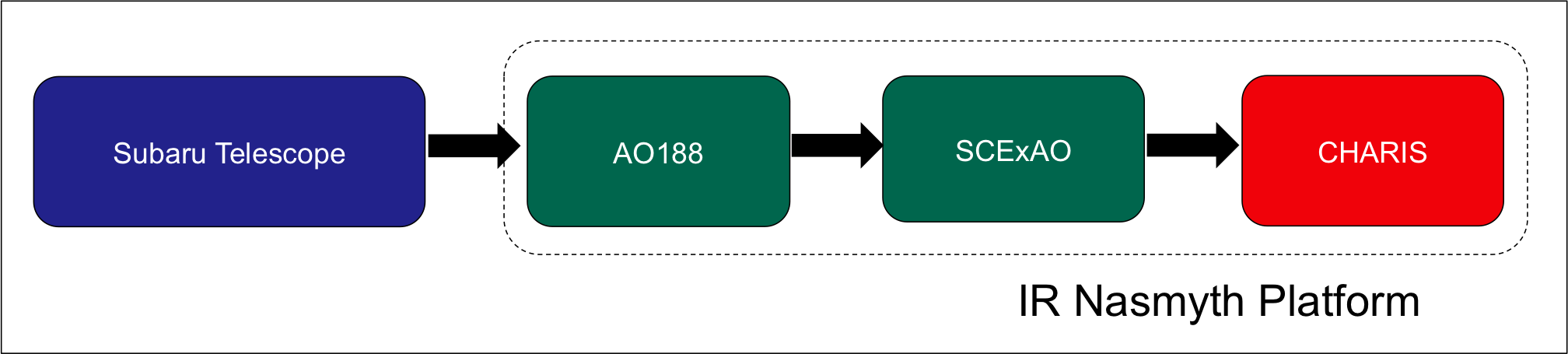}
\caption[example] 
{ \label{fig:nasmyth} 
Conceptual layout of the Subaru Nasmyth platform as it will function with CHARIS.  Subaru will feed its beam to AO188, which will perform a low-order wavefront correction.  SCExAO will perform a high-order correction and add coronagraphy, delivering a very high-quality beam to CHARIS.
}
\end{figure}

CHARIS is the successor to the HiCIAO high-contrast camera\cite{Hodapp+Suzuki+Tamura+etal_2008} on the Subaru Telescope.  HiCIAO received its beam from AO188\cite{Hayano+Takami+Oya+etal_2010}, a 188-actuator adaptive optics system.  AO188 is currently being augmented by SCExAO, which uses a pyramid wavefront sensor, active speckle nulling\cite{Martinache+Guyon+Jovanovic+etal_2014}, and a choice of several coronagraphs to dramatically improve contrast.  Figure \ref{fig:nasmyth} shows a schematic of the Nasmyth bench as it will function with CHARIS.

\begin{table}
\centering
\caption{CHARIS Specifications \label{tab:charis_spec}}
\smallskip
\begin{tabular}{ l c r}
Spaxel Scale & \multicolumn{2}{r}{$0.\!\!''015$ ($\frac{1}{2}\lambda$/D at 1.15 $\rm \mu m$)} \\
Spectral Range & \multicolumn{2}{r}{1.15 -- 2.4 $\rm \mu m$}\\
Observing Mode & low-R & high-R \\
Spectral Resolution & $\sim$18 & $\sim$75 \\
Bandwidth & 1.2 $\rm \mu m$ & 0.3 $\rm \mu m$\\
Detector & \multicolumn{2}{r}{Hawaii 2RG}\\
Field of View & \multicolumn{2}{r}{$2.\!\!''07 \times 2.\!\!''07$} 
\end{tabular}
\end{table}

\section{Current High-Contrast Imaging on Subaru: the SEEDS Survey}

The Strategic Exploration of Exoplanets and Disks with Subaru (SEEDS) Survey\cite{Tamura_2009} is a five year, 120 night strategic observing program on the Subaru Telescope, currently nearing completion.  SEEDS uses HiCIAO, the predecessor of CHARIS, to search for exoplanets and characterize dust disks using high-contrast imaging.  SEEDS consists of two main observing modes:
\begin{enumerate}
\item Polarized differential imaging (PDI), in which a Wollaston prism separates the two linear polarization directions.  These polarizations are subtracted to construct a map of polarized intensity.
\item Angular differential imaging (ADI), in which field rotation is used to distinguish point sources from quasistatic speckles.  
\end{enumerate}
PDI has been used very successfully to image young circumstellar disks.  Using PDI, SEEDS observations have discovered large central gaps\cite{Hashimoto+Tamura+Muto+etal_2011, Hashimoto+Dong+Kudo+etal_2012} and spiral structures\cite{Muto+Grady+Hashimoto+etal_2012} in disks, which may point to unseen planets.

Most of SEEDS' time has been spent searching for substellar companions using ADI observations of several hundred nearby stars\cite{}.  HiCIAO discovered one brown dwarf, GJ 758 B\cite{Thalmann+Carson+Janson+etal_2009}, in commissioning, and an additional two substellar companions, GJ 504 b\cite{Kuzuhara+Tamura+Kudo+etal_2013} and $\kappa$ And b\cite{Carson+Thalmann+Janson+etal_2013}, during the main SEEDS survey.  GJ 504 b is a particularly interesting object, with a low mass ($\sim$3--8 $M_{\rm Jup}$), a low temperature, and strong methane absorption\cite{Janson+Brandt+Kuzuhara+etal_2013}, though both it and GJ 758 B lie outside CHARIS's field-of-view.  The substellar companion to $\kappa$ And has a very large uncertainty in its mass, from $\sim$12--50 $M_{\rm Jup}$\cite{Carson+Thalmann+Janson+etal_2013, Hinkley+Pueyo+Faherty+etal_2013}, due to controversy over the host star's age.  CHARIS observations could provide spectral signatures of $\kappa$ And b's surface gravity, helping to resolve this controversy (see Section \ref{sec:characterization}).

SEEDS observations, in addition to discovering new substellar companions, have helped place stringent constraints on the population of wide-separation exoplanets\cite{Brandt+McElwain+Turner+etal_2014}.  SEEDS data, combined with other surveys, require that the population of planets found by radial velocity surveys cannot be extrapolated beyond $\sim$30--70 AU (depending on the planetary atmosphere model).  The data are also most consistent with a single substellar population, from $\sim$5 $M_{\rm Jup}$ up to the hydrogen burning limit, accounting for the SEEDS companions.  The following sections describe the implications of these findings for CHARIS.

\section{Substellar Populations}

CHARIS will serve as both a survey/discovery instrument and a characterization tool.  The population of exoplanets and brown dwarfs available to discovery depends on the mass function, substellar cooling curves, and the age distribution of the target sample.  Radial velocity surveys have extensively characterized the population of objects $\lesssim$10 $M_{\rm Jup}$ within $\sim$3 AU\cite{Cumming+Butler+Marcy+etal_2008}, while direct-imaging survey results seem to be mostly consistent with a higher-mass, brown dwarf-like distribution from $\sim$5--70 $M_{\rm Jup}$ at separations of up to $\sim$1000 AU \cite{Brandt+McElwain+Turner+etal_2014}.  

\begin{figure}
\centering
\includegraphics[width=0.512\linewidth]{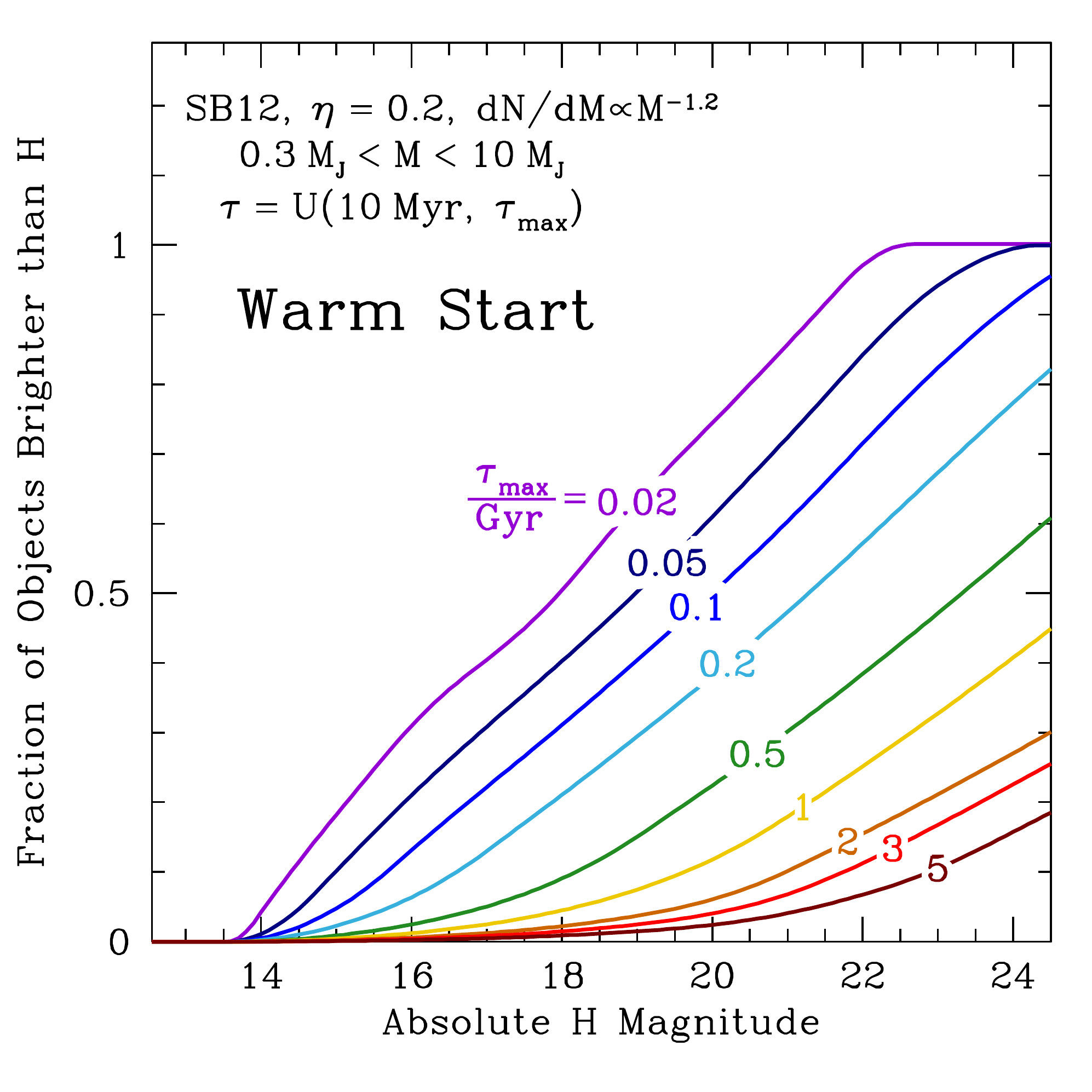}
\includegraphics[width=0.448\linewidth]{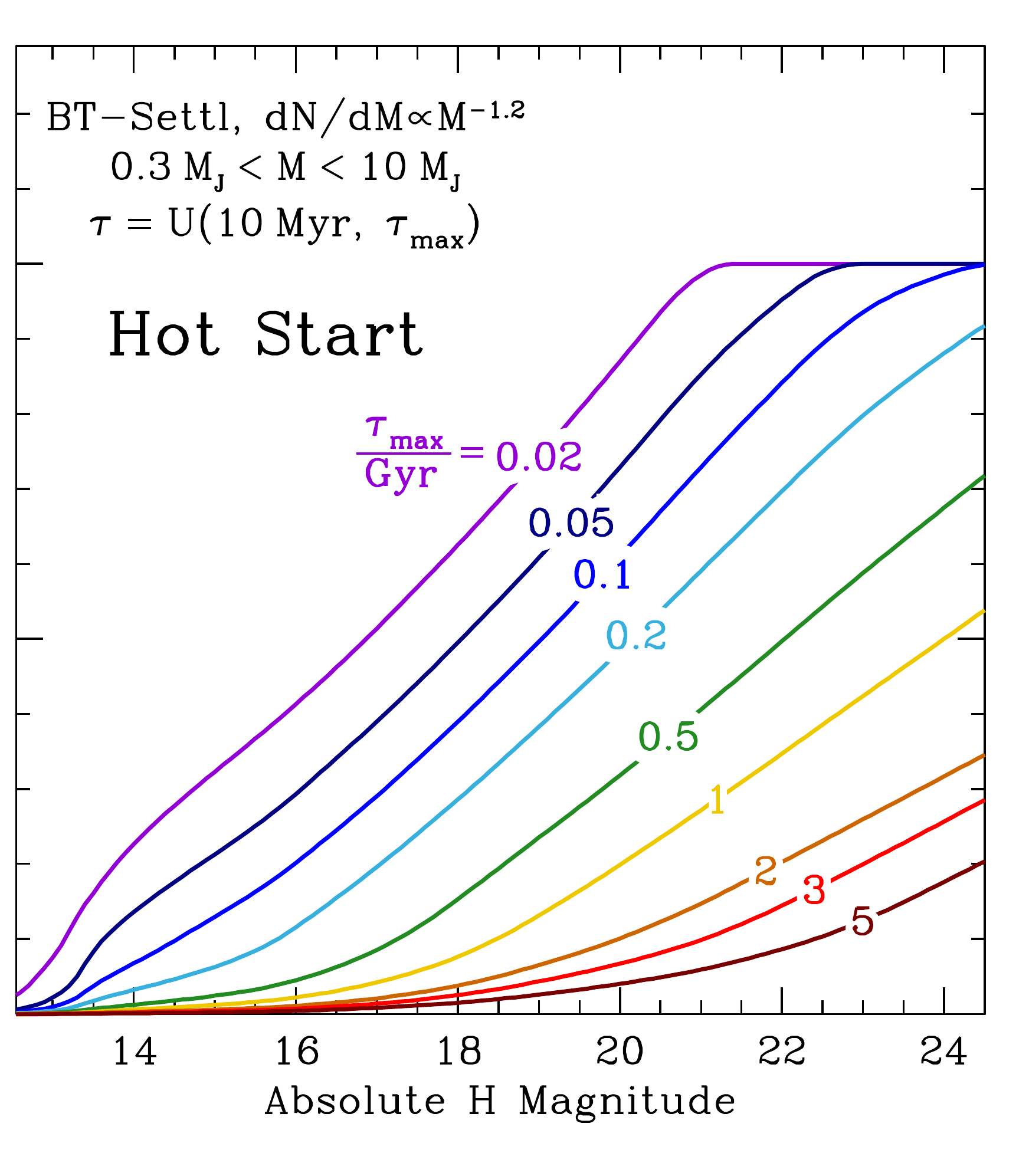}
\caption{Luminosity functions for the mass distribution derived from radial velocity surveys\cite{Cumming+Butler+Marcy+etal_2008} for planets within $\sim$3 AU and between 0.3 and 10 $M_{\rm Jup}$.
\label{fig:planet_lf}
}
\end{figure}

\begin{figure}
\centering
\includegraphics[width=0.512\linewidth]{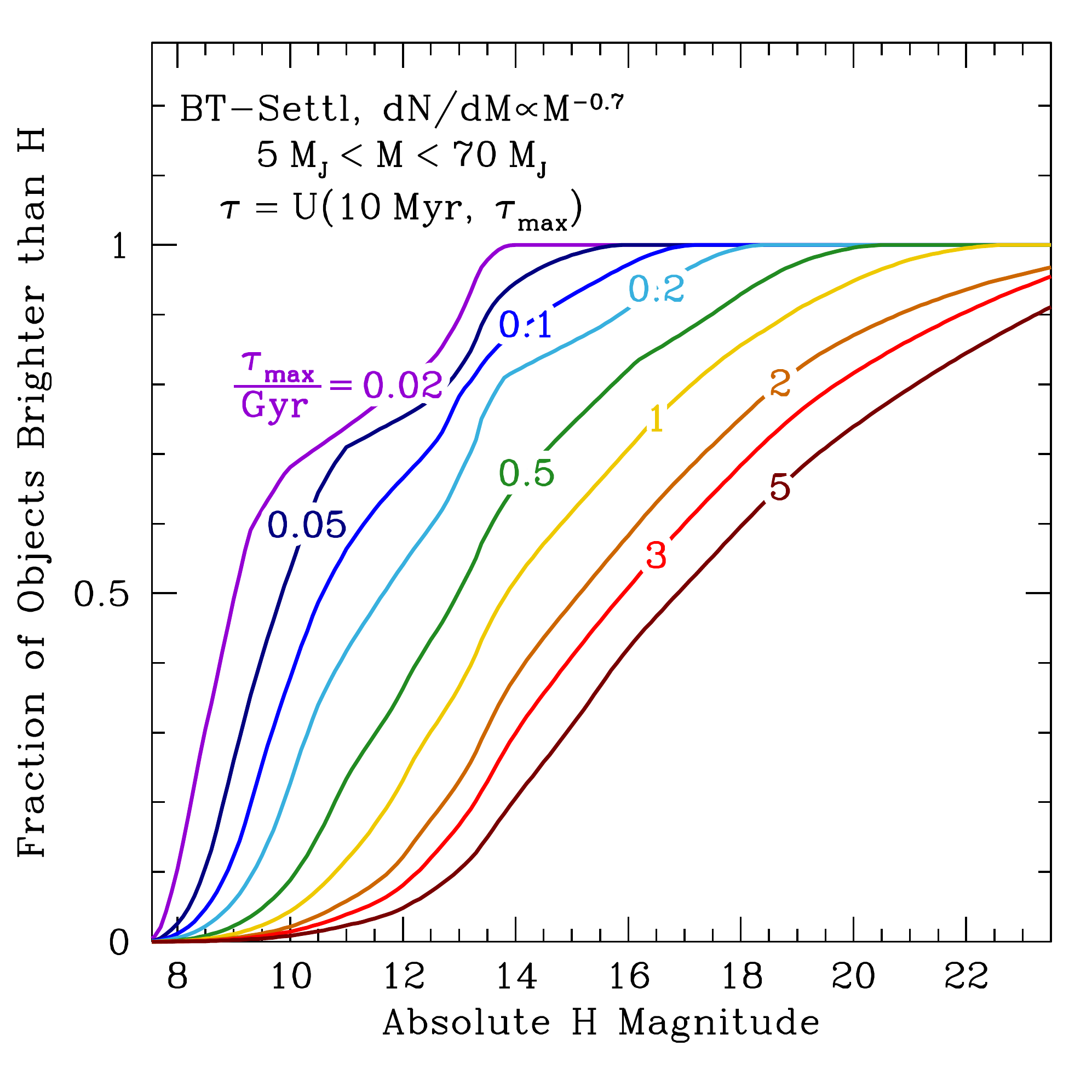}
\caption{Luminosity function for the mass distribution derived for directly imaged substellar companions at separations of tens to hundreds of AU\cite{Brandt+McElwain+Turner+etal_2014}.
\label{fig:bd_lf}
}
\end{figure}

Given substellar cooling models and measured mass and age distributions, we can compute substellar luminosity functions.  These are shown in Figure \ref{fig:planet_lf} for the radial velocity mass function\cite{Cumming+Butler+Marcy+etal_2008} with both the BT-Settl hot-start cooling models\cite{Allard+Homeier+Freytag_2011} and Spiegel/Burrows warm-start models\cite{Spiegel+Burrows_2012}.  The warm-start models have initial entropies just slightly higher than those of the coldest Spiegel/Burrows models.  The luminosity functions are cumulative distribution functions, showing the fraction of objects more luminous than a given absolute magnitude.  All curves assume a uniform distribution of ages between 10 Myr and the age indicated.  
Figure \ref{fig:bd_lf} shows the same luminosity functions as Figure \ref{fig:planet_lf}, but for the substellar mass function derived from direct imaging surveys\cite{Brandt+McElwain+Turner+etal_2014}.

As Figures \ref{fig:planet_lf} and \ref{fig:bd_lf} indicate, the best targets for direct imaging planet searches are nearby young stars.  Unfortunately, such stars are rare.  The nearest starforming regions, such as Taurus and the $\rho$ Ophiuchus cloud complex, lie at distances of $\sim$100--150 pc\cite{}.  Several moving groups, with ages ranging from $\sim$10--100 Myr, lie at distances of $\sim$20--50 pc\cite{Malo+Doyon+Lafreniere+etal_2013}.  However, these groups collectively contain just a few hundred stars, nearly all of which have already been targeted by one or more high-contrast instruments.  Nearby field stars can be much closer but are typically several Gyr old and difficult to date.  A CHARIS survey thus has three basic target selection strategies:
\begin{enumerate}
\item very young but relatively distant stars ($\sim$5--10 Myr, $\sim$100 pc);
\item young and relatively nearby stars which have mostly been observed already with previous-generation instruments ($\sim$10--100 Myr, $\sim$20--50 pc); or 
\item the very nearest stars ($\sim$1--10 Gyr, $\sim$2--20 pc).  
\end{enumerate}
The available discovery space, in projected separation and contrast, depends on distance and previous observations, while the discovery space in companion mass also depends on target age.  Figures \ref{fig:planet_lf} and \ref{fig:bd_lf} show the luminosity distributions of these target samples, both for planet-like and brown dwarf-like distributions.  The violet curves correspond to the age distribution in very young clusters, while the blue curve corresponds to that for young moving groups, and the dark red curves show the approximate luminosity functions for the local field.  Each population has a typical distance $\sim$2.5 times that of the next closest, corresponding to a distance modulus (difference between apparent and absolute magnitudes) of $\sim$2 magnitudes.  

\section{Sensitivity}

CHARIS's low-dispersion mode will allow for SDI over an unprecedented wavelength range, and should vastly improve the achievable contrast at small angular separations.  SDI uses the fact that speckles, which are diffraction spots in the stellar PSF, scale with wavelength in their separation from the star's location.  A speckle at separation $r$ is thus spread over a range $\delta r = r \delta \lambda/\lambda$, where $\delta \lambda/\lambda$ is the filter width.  For an instrument like GPI, operating on a single $\sim$20\% band at a time (e.g.~$H$, from $\sim$1.5--1.8 $\mu$m), $\delta r \approx 0.2 r$.  

SDI is only effective at separations where the radial extent of a speckle is large compared to the size of the planet's PSF.  Thus, for a 20\% bandpass, SDI requires an angular separation $\gtrsim$$1/0.2=5$ $\lambda/D$, with its effectiveness dropping sharply near this separation.  For a wavelength at the center of the bandpass, the furthest reference wavelength is just $\delta\lambda/2$ away.  As a result, SDI will only be effective for all wavelengths at separations $\gtrsim$10 $\lambda/D$, or $\gtrsim$$0.\!\!''4$ for an 8 meter telescope at 1.6 $\mu$m.  CHARIS, with a full factor of 2 in simultaneous wavelength coverage, will be able to fully exploit SDI down to separations of just a few $\lambda/D$, $\sim$$0.\!\!''1$.

CHARIS will sit behind SCExAO on Subaru's Nasmyth platform, and will benefit from a high-quality input beam.  SCExAO expects to deliver contrast approaching $10^{-5}$ up to an inner working angle of $\sim$0$.\!\!''1$, $\sim$3$\lambda/D$ at the $J$ band.  CHARIS will also be background limited in its low-dispersion mode, allowing it to reach a limiting $H$-band magnitude of $\sim$24.  

SCExAO is currently in its final stages of commissioning.  Though the pyramid wavefront sensor is now installed, the SCExAO software has not yet reached the full number of spatial modes it will ultimately correct.  We therefore cannot use on-sky measurements of the AO performance, and instead use a heuristic fitting function that mimics the basic performance achieved by the GPI team\cite{Macintosh+Graham+Ingraham+etal_2014}, but with better contrast at very small angular separations.  The latter assumption is due both to SCExAO's expected contrast at a $\sim$$0.\!\!''1$ inner working angle with the PIAA coronagraph and to our exceptionally broad wavelength coverage in CHARIS's $R \sim 18$ dispersion mode, enabling SDI to be effective even at these small angular separations.  

In our contrast-limited case, we provisionally adopt the equation
\begin{equation}
\Delta m = 19 - 8\exp\left[-\frac{\delta \phi}{0.\!\!''8} \right]~,
\label{eq:contrast}
\end{equation}
with $\delta \phi$ being the angular separation and $\Delta m$ the contrast in magnitudes.  Equation \eqref{eq:contrast} gives a contrast of up to 12 magnitudes $1.6 \times 10^{-5}$ at $0.\!\!''1$ and 14.7 magnitudes at $0.\!\!''5$.  This is significantly better than GPI's contrast at very small separations, and very slightly better (by less than 1 magnitude) at $0.\!\!''5$.\cite{Macintosh+Graham+Ingraham+etal_2014}  We combine Equation \eqref{eq:contrast} in quadrature with a flat magnitude limit of 23 to obtain effective contrast curves.

\section{CHARIS Discovery Prospects}

A multitude of recent surveys have imaged $\sim$1000 young, nearby stars to search for faint substellar companions.  The additional discovery space CHARIS will provide will be at larger contrasts, but especially at smaller separations.  Archival high-contrast imaging is already sensitive to most companions above the deuterium burning limit (i.e.~with masses $\gtrsim$13 $M_{\rm Jup}$) at separations $\gtrsim$10 AU.  CHARIS will fill in the space of companions down to a few $M_{\rm Jup}$, as low as $\sim$1 $M_{\rm Jup}$ around the very youngest and closest stars, at separations of a few AU.

\begin{figure}
\centering
\includegraphics[height=0.48\linewidth]{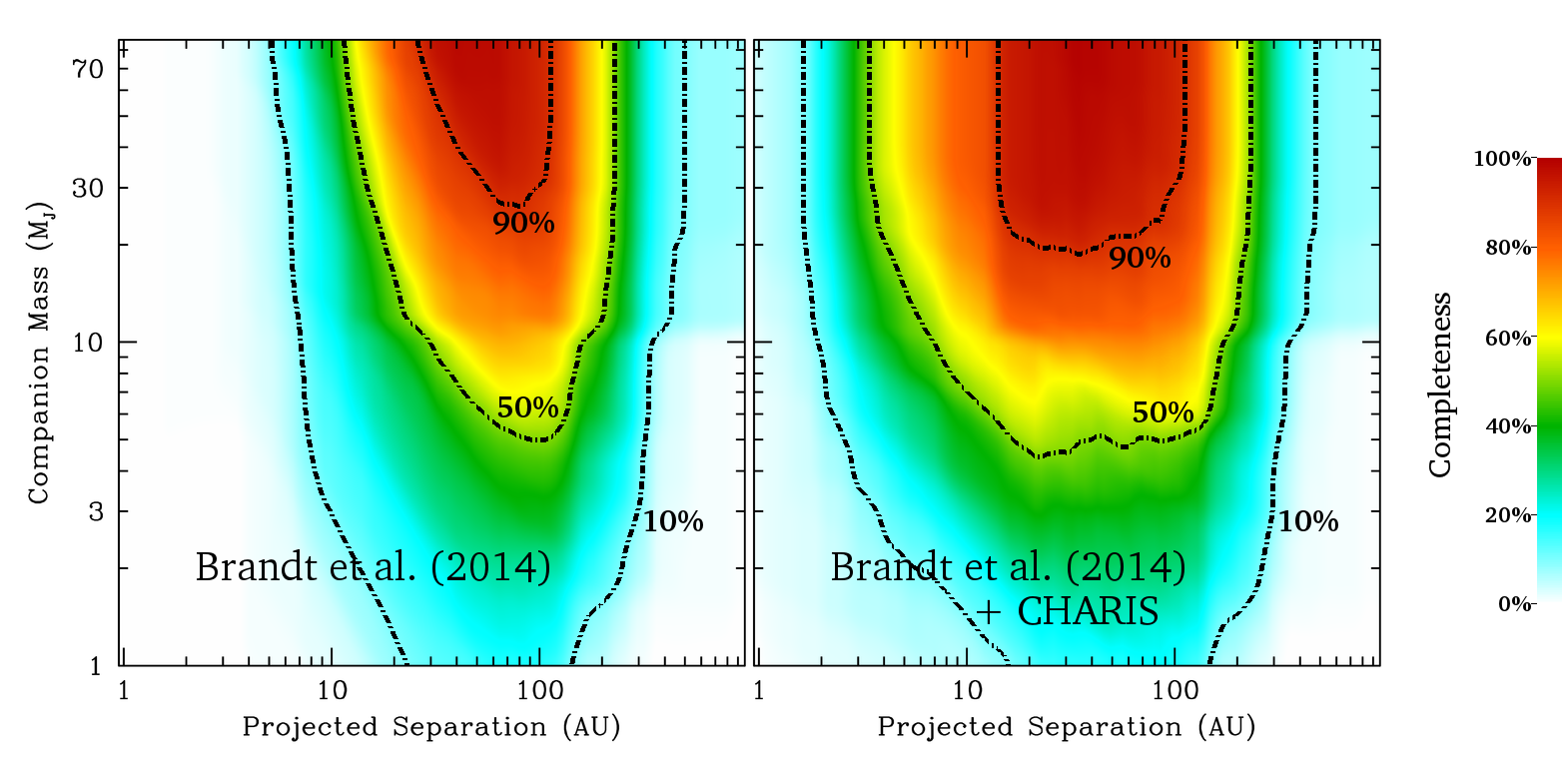}
\caption{Left panel: completeness of a high-contrast sample of $\sim$250 stars formed by combining several surveys\cite{Brandt+Kuzuhara+McElwain+etal_2014, Janson+Brandt+Moro-Martin+etal_2013, Yamamoto+Matsuo+Shibai+etal_2013, Lafreniere+Doyon+Marois+etal_2007, Biller+Liu+Wahhaj+etal_2013}
and assuming the BT-Settl hot-start models\cite{Allard+Homeier+Freytag_2011}.  
Right panel: the expected completeness of the same sample of $\sim$250 stars after observing those visible from the Northern Hemisphere ($\sim$200 of the $\sim$250) with CHARIS.
\label{fig:completeness_comp}
}
\end{figure}

The left panel of Figure \ref{fig:completeness_comp} shows the completeness of a heterogeneous sample of $\sim$250 nearby stars that was recently compiled from several different surveys\cite{Brandt+Kuzuhara+McElwain+etal_2014, Janson+Brandt+Moro-Martin+etal_2013, Yamamoto+Matsuo+Shibai+etal_2013, Lafreniere+Doyon+Marois+etal_2007, Biller+Liu+Wahhaj+etal_2013} for a statistical analysis\cite{Brandt+McElwain+Turner+etal_2014}.  The stars range in spectral type from mid M to early A, and in age from $\sim$10 Myr to several Gyr.  The large spread in ages arises largely because of the difficulty of dating individual stars.  The most reliable ages come from membership in nearby moving groups, which is often difficult to assign.  However, recent Bayesian analyses\cite{Malo+Doyon+Lafreniere+etal_2013, Gagne+Lafreniere+Doyon+etal_2014} are beginning to treat moving group membership in a more systematic way, giving membership probabilities for any given star based on its position in phase space.  The figure assumes the BT-Settl hot-start models\cite{Allard+Homeier+Freytag_2011} for all substellar objects.  This is likely to be a very good assumption for objects formed by a direct gravitational collapse, though those that formed more slowly, by accretion onto a rocky core, could radiate away much of their initial heat during formation\cite{Marley+Fortney+Hubickj+etal_2007}.

The most straightforward approach to a survey with CHARIS would be to observe the nearest and youngest stars, which have almost invariably been included in other high-contrast imaging data sets.  We provide a first estimate of the possible yield of this approach by taking the same sample of $\sim$250 stars, and improving the existing contrast curves at small angular separations for those visible from the Northern Hemisphere ($\sim$200 of our $\sim$250).  In practice, we would prioritize the targets, selecting fewer than 200 stars from this sample, and choosing additional targets from recent analyses of nearby moving groups.  The right panel of Figure \ref{fig:completeness_comp} shows the estimated completeness of the same sample of $\sim$250 stars after adding CHARIS observations, assuming a survey of all $\sim$200 potential targets in the sample.

\begin{figure}
\centering
\includegraphics[height=0.48\linewidth]{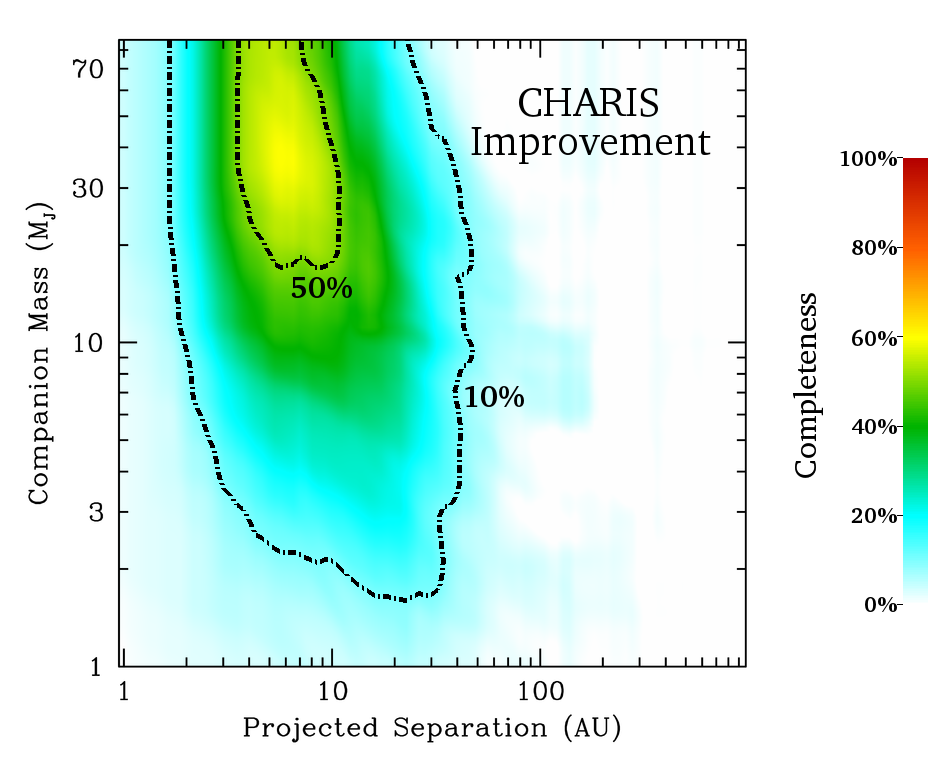}
\caption{The additional completeness provided in a merged sample of $\sim$250 stars\cite{Brandt+McElwain+Turner+etal_2014} by observing the $\sim$200 targets visible from the Northern Hemisphere with CHARIS.  This is the difference of the two completeness distributions shown in Figure \ref{fig:completeness_comp}.  CHARIS would dramatically improve the completeness to brown dwarfs at Solar system-scale separations, and, for the younger stars in our sample, would be sensitive to planets of a few Jupiter masses at a few AU.  The completeness at low masses and small separations would be much higher if we restricted the figure to truly young stars visible from the Northern Hemisphere, less than half of the full sample of $\sim$250 stars.  
\label{fig:completeness_impr}
}
\end{figure}

Figure \ref{fig:completeness_impr} isolates the additional sensitivity provided by CHARIS, the difference between the left and right panels of Figure \ref{fig:completeness_comp}.  For companions in the brown-dwarf mass regime at separations comparable to those of the giant planets in our own Solar system, CHARIS could increase the completeness on the full sample by a full $\sim$50\%, from $\sim$30\% to more than 80\%.  This would provide nearly a complete census of brown dwarfs on outer Solar system scales.  For the youngest and nearest stars, CHARIS would also dramatically expand our sample's completeness to exoplanets of several $M_{\rm Jup}$, again at separations of a few AU, comparable to the scale of the outer Solar system, assuming that the hot-start BT-Settl models accurately predict their evolution.  Figure \ref{fig:completeness_impr} would have much higher completeness levels below the deuterium burning limit if we restricted the analysis to truly young stars visible from the Northern Hemisphere, less than half of the full sample of $\sim$250 stars.

The discovery yield of such an observing program depends on the completeness shown in Figure \ref{fig:completeness_impr} and on the distribution function of the substellar population.  A recent statistical analysis\cite{Brandt+McElwain+Turner+etal_2014} found that, assuming hot-start models, a distribution derived from radial-velocity observations, $dN/dMda \propto M^{-1.3}a^{-0.6}$ with $\sim$10\% of stars having companions between 0.3 and 10 $M_{\rm Jup}$ and separations from 0.03 to 3 AU, cannot be extrapolated past $\sim$30 AU.  If we assume this distribution to truncate as a step function at 20 AU, then Figure \ref{fig:completeness_impr} predicts $\sim$1--3 new discoveries.  This number could be significantly higher with a larger truncation radius, an upturn in planet frequency at $\sim$5--10 AU, or with a more careful choice of target stars.  On the other hand, it could be overly optmistic if the BT-Settl models over-predict the luminosity of young planets and/or the distribution function falls off far interior to 20 AU.

For a brown-dwarf-like distribution\cite{Brandt+McElwain+Turner+etal_2014}, $dN/dMda \propto M^{-0.7} a^{-0.8}$ and normalized so that $\sim$5\% of stars have companions between 5 and 70 $M_{\rm Jup}$ and between 10 and 100 AU, the predicted discovery yield from Figure \ref{fig:completeness_impr} is $\sim$1 object.  A much better approach to find these higher-mass companions would be to target the very nearest stars without archival high-contrast observations.  CHARIS's contrast and inner working angle would render objects above the deuterium burning threshold detectable even after several Gyr of cooling.

\section{Spectroscopy and Characterization}\label{sec:characterization}

In addition to its unique $R \sim 18$ discovery mode, CHARIS will offer a higher resolution, $R \sim 75$ mode.  The dispersion in this mode is remarkably constant as a function of wavelength, enabling CHARIS to cover an entire band, either $J$, $H$, or $K$, with its $\sim$15 spectral measurements.  CHARIS will use this mode to take low-resolution spectra of nearby exoplanets and brown dwarfs.

\begin{figure}
\centering
\includegraphics[width=0.6\linewidth]{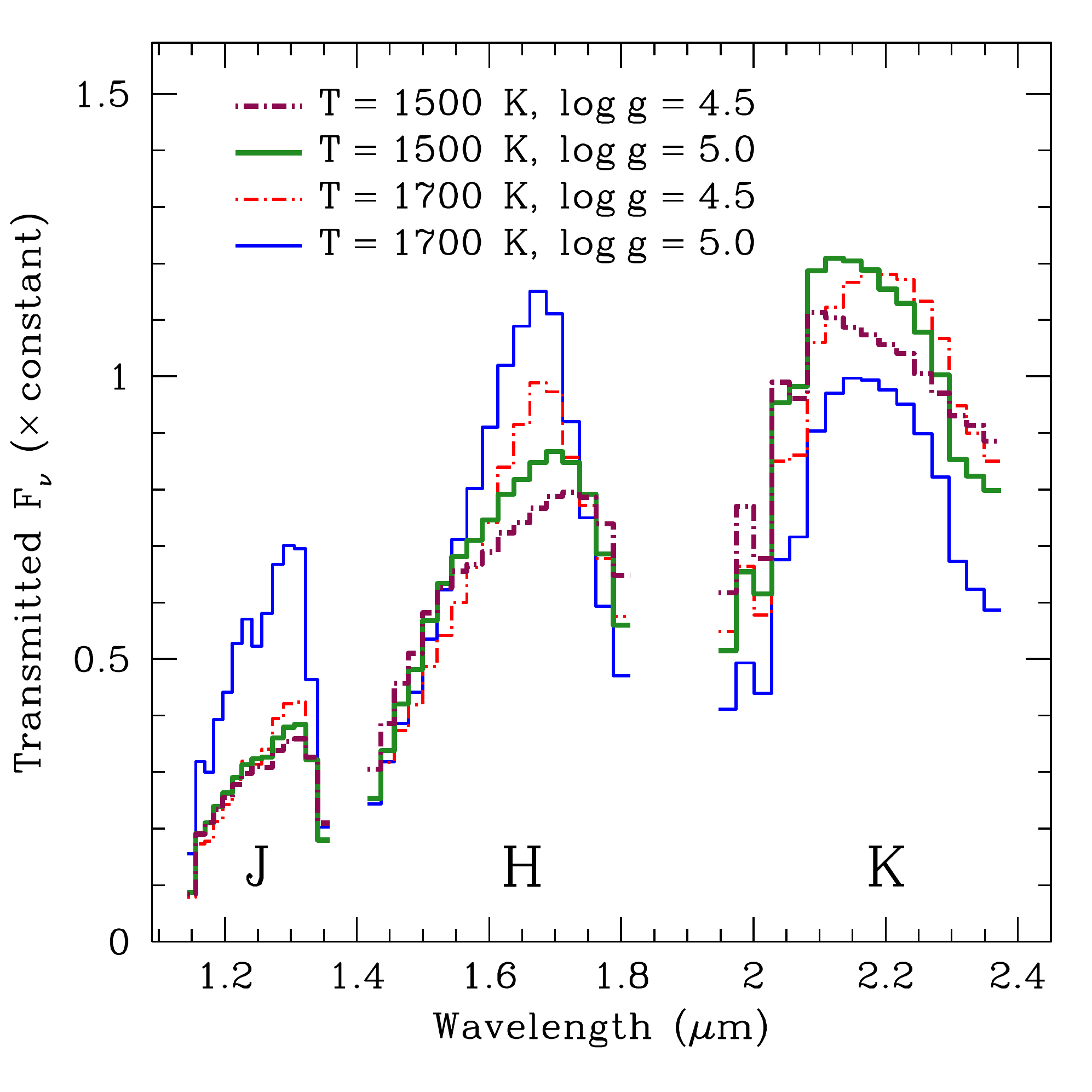}
\caption{Example of exoplanet characterization with CHARIS.  The spectra shown are BT-Settl models with two different gravities and effective temperatures, corresponding roughly to the temperature inferred for $\kappa$ And b, and to the effective gravities it would have as either a $\sim$13 $M_{\rm Jup}$ young object or an older, $\sim$50 $M_{\rm Jup}$ brown dwarf.  }
\label{fig:kapand}
\end{figure}

Figure \ref{fig:kapand} shows the value of these measurements, using $\kappa$ And b as an example.  Its host star, $\kappa$ And, is a late B/early A star with a possible, but uncertain, membership in the Columba moving group\cite{Gagne+Lafreniere+Doyon+etal_2014}.  Should $\kappa$ And be a moving group member, it would have an age $\sim$30 Myr, which would imply a companion mass of $\sim$13 $M_{\rm Jup}$ (and $\log g \sim 4.2$) given its infrared luminosity\cite{Carson+Thalmann+Janson+etal_2013}.  However, the star could be much older, which would imply a companion mass closer to 50 $M_{\rm Jup}$, and $\log g \sim 5.0$\cite{Hinkley+Pueyo+Faherty+etal_2013}.  Low-resolution spectroscopy can break this degeneracy, by measuring spectral shapes in the $H$ and $K$ bands that are sensitive to the surface gravity.  

\section{Summary} 

CHARIS, scheduled for delivery to the Subaru Telescope in early 2016, will be the first high-contrast IFS on an 8m telescope in the Northern Hemisphere.  CHARIS will offer a low-resolution dispersion mode, unique among this class of instruments, that will enable it to simultaneously take spectra across the near-infrared, from 1.15--2.4 $\mu$m.  This dispersion mode will combine with a very high-quality coronagraphic beam from SCExAO, Subaru's new extreme adaptive optics system, to enable contrasts of $\sim$$10^{-5}$ at an inner working angle of just a few $\lambda/D$, $\sim$$0.\!\!''1$.  CHARIS will also offer a higher resolution mode with $R \sim 75$, in which it can take spectra of a single near-infrared band, $J$ or $H$ or $K$, at a time.

CHARIS will offer exceptional sensitivity to exoplanets and brown dwarfs at small angular separation.  We have estimated the yield of a CHARIS survey consisting of repeat observations of Northern Hemisphere stars targeted by high-contrast cameras, finding that such a survey of $\sim$200 stars might detect $\sim$1--3 planets.  This sample contains many older stars; a careful optimization of the targets could increase the yield by a factor of a few.  The yield depends sensitively on the details of the exoplanet distribution function and on the substellar cooling model adopted.  A favorable distribution function, extending to $\gtrsim$20 AU with an upturn at $\sim$10 AU, could significantly increase the number of detectable planets.

CHARIS will also help to characterize exoplanets and brown dwarfs, including those discovered by surveys like SEEDS, a large observing program using the Subaru Telescope's current AO and high-contrast camera.  CHARIS spectra, for example, can help resolve whether $\kappa$ And b is a young $\sim$12 $M_{\rm Jup}$ object or an older $\sim$50 $M_{\rm Jup}$ brown dwarf by looking for spectral signatures of low surface gravity.  CHARIS will also combine with similar instruments including GPI and SPHERE in the Southern Hemisphere, to build a library of substellar spectra extending to masses well below the deuterium burning limit of $\sim$13 $M_{\rm Jup}$.

\acknowledgments       
 
This work was performed with the support of 
the Japanese government's Ministry of Education, Culture,
Sports, Science and Technology through grant-in-aid number
23103002 of the program for Scientific
Research on Innovative Areas.  TDB gratefully acknowledges support from the Corning Glass Works Foundation through a fellowship at the Institute for Advanced Study.

\bibliographystyle{spiebib}
\bibliography{refs}

\end{document}